# Multiple Narrow-band signals Direction Finding with TMLA by Nonuniform Period Modulation

Kebin Liu[(1)], Lening Zhang[(1)], Qingkui Zhan[(1)] and Chong He[(1)]
Dept. of Electronic Engineering Shanghai Jiao Tong University Shanghai, China (1174587701@qq.com)

*Abstract*—A new array signal reconstruction and signal-channel DOA estimation method based on TMLA by nonuniform period modulation is proposed. By using non-uniform period modulation, the harmonic component produced by different elements could be separated. Therefore, the conventional snapshot could be reconstructed by analyzing the spectrum of the combined signal. Then spatial spectrum estimation method is used to implement DOA estimation. Numerical simulations are provided to verify the feasibility and accuracy of the proposed method. Since the duration of the signal in the frequency domain analysis processed in a single time is very short, this method is also applicable to narrowband signals. Another highlight is that this method can simultaneously measure the number of the elements-1 angle of incident signals.

## I. INTRODUCTION

Time-modulated arrays add a dimension in time by introducing periodic switches on the array to periodically control the array elements.[1] The origin of time-modulated arrays dates to the 1950s, and TMA-related research has focused on beam synthesis, adaptive beamforming, sideband suppression, and direction-finding. By introducing a new time dimension to the array, TMA can effectively reduce the dynamic range of the amplitude excitation in the feeding network. With the development of TMA theory, the unique harmonic characteristics due to the modulation have been applied to power distribution systems, array calibration, and direction-finding. [2]

Direction-finding using TMA has received more and more attention in recent years because of its unique features, such as requiring only a single channel and generating characteristic harmonic features after periodic modulation. In this proposed method, the received signals on each array element are modulated using a designed function at first and then combined. Finally, frequency domain analysis is performed on the combined signals. Since the frequencies of the modulation functions on each array element are different, the received signals on different array elements can be distinguished by using different frequencies after the analysis of the combined signal. In addition, this method also makes special requirements on the modulation frequency, so that the duration of the signal for a single frequency domain analysis can be very short, once the bandwidth of the baseband signal is narrow enough, it can be considered as constant in single frequency domain analysis. For such an incoming signal it is possible to use the present method for direction finding. This greatly relaxes the previous requirement of using harmonic features for direction finding. After reproducing the conventional snapshot in the frequency domain, the conventional spatial spectrum estimation algorithm is then used[3], so that multiple incoming signals can be oriented. This is also a considerable improvement over the previous method.

## II. THEORY

Consider an N-element linear TMA with element spacing D and excited by $M$ narrow-band signals from the far-field with the carrier frequency $f_c$ and bandwidth $B_m, m = 1, 2, ..., M$ are impinging on the antenna array.

For the first element, the $m^{th}$ incoming signal is

$$s_m(t) = s_{b(m)}(t)e^{j2\pi f_c t} \quad (1)$$

Where $s_{b(m)}(t)$ is the baseband signal of the $m^{th}$ incoming signal. The receiving signal $x_1(t)$ the first antenna element can be written as

$$x_1(t) = \sum_{m=1}^{M} s_{b(m)}(t)e^{j2\pi f_c t} \quad (2)$$

The receiving signal $x_n(t)$ in the $n^{th}$ antenna element can be written as

$$x_n(t) = \sum_{m=1}^{M} s_{b(m)}(t-(n-1)\tau_m)e^{-j(n-1)\beta D \sin\theta_m} \cdot e^{j2\pi f_c t} \quad (3)$$

$\tau_m$ represents the time duration required for the $m^{th}$ signal to pass through the distance between elements.

$$\tau_m = D \cdot \sin\theta_m / c \quad (4)$$

$\theta_m$ is the direction of arrival of the $m^{th}$ incoming signal and $\beta = 2\pi/\lambda$ is the wavenumber, and $\lambda$ is the carrier wavelength. To avoid direction-finding ambiguity, the following condition should be satisfied

$$\beta D \sin\theta \leqslant \pi \quad (5)$$

Since the incident angle is unknown, D needs to consider all possible cases. It is obvious that $sin\theta_{maximum} = 1$ and increasing the length of baseline as much as possible without bringing ambiguity can improve the directional accuracy. [6] Therefore, we normally set $D = \lambda/2$. Then the TMLA with nonuniform period modulation is used. The modulation period added to each element varies with each other. The modulation period of the reference element is $T_p$. Based on this, the modulation frequency of the $n^{th}$ element is [4]

$$f_{p,n} = 2^{n-1} f_p \quad (6)$$

When all incoming signals are sinusoidal signals with the same frequency, the frequency of harmonics produced by different element is different. The property will be used for multiple narrow-band signals direction finding.

The modulation function added to the $n^{th}$ element is

$$U_n(t) = \begin{cases} 1, & iT_{p,n} \leqslant t < iT_{p,n} + \frac{1}{2}T_{p,n} \quad i \text{ is interger} \\ -1, & iT_{p,n} + \frac{1}{2}T_{p,n} \leqslant t < iT_{p,n} + T_{p,n} \end{cases} \quad (7)$$

Decomposition $U_n(t)$ using the Fourier series formula, the signal can be represented as follow

$$U_n(t) = \sum_{k=-\infty}^{+\infty} \frac{j}{k\pi}(e^{-jk\pi}-1)e^{jk2\pi \cdot 2^{n-1}f_p t} \quad (8)$$

For the $n^{th}$ element, the received signal after modulation is

$$\sum_{k=-\infty}^{+\infty}\left\{\sum_{m=1}^{M} s_{b(m)}(t-(n-1)\tau_m)e^{-j(n-1)\beta D\sin\theta_m}\right\}\alpha_{n,k} \cdot e^{j2\pi f_c t}e^{jk2\pi f_{p,n} t}$$

$\alpha_{n,k}$ is the Fourier coefficient. Then after the combiner

$$\sum_{n=1}^{N}\left\{\sum_{k=-\infty}^{+\infty}\left\{\sum_{m=1}^{M} s_{b(m)}(t-(n-1)\tau_m)e^{-j(n-1)\beta D\sin\theta_m}\right\}\alpha_k\right\} \cdot e^{j2\pi f_c t}e^{jk2\pi \cdot 2^{n-1}f_p t} \quad (9)$$

Since all incoming signals are narrowband signals, if $f_p \gg B_m$ is small enough, it can be assumed that all baseband signals are constant within $T_p$. The mathematical expression is written as:

$$S_{b(m)}(t) \approx S_{b(m)}(t+T_c) \approx S_{b(m)}(t-(N-1)\tau_m) \approx S_{b(m)}(t-(N-1)\tau_m+T_c) \quad (10)$$

where the first equation indicates that the instantaneous value of the baseband signal of the $m^{th}$ incoming signal received on the first element at time $t$ is approximately equal to the instantaneous value at $t + T_c$ on the first element, the second equation indicates that the instantaneous value at time $t + T_c$ on the first element is approximately equal to the instantaneous value at $t$ on the $N^{th}$ array element and the third equation indicates that the instantaneous value at $t$ on the $N^{th}$ element is approximately equal to the instantaneous value at time $t + T_c$ on the $N^{th}$ array element instantaneous value on the $N^{th}$ array element. It can be summarized that the baseband signal received on either array element can be considered a constant when sampling is performed within $T_p$. The baseband signal is denoted by $m_1, m_2, \ldots, m_M$. Therefore, sampling this signal could be regarded as sampling the signal below

$$\sum_{n=1}^{N}\left\{\sum_{k=-\infty}^{+\infty}\left\{\sum_{m=1}^{M} m_m e^{-j(n-1)\beta D\sin\theta_m}\right\}\alpha_k\right\} \cdot e^{j2\pi f_c t}e^{jk2\pi \cdot 2^{n-1}f_p t} \quad (11)$$

Based on the Fourier series transform formula and the limitation of narrowband signal bandwidth The proposed method normally requires 1. $T_p = kT_c$ 2. $f_p \gg B_m$. $T_p$ needs to be adjusted according to the $f_c$ of the received signal. According to the Fourier series formula, the accurate frequency spectrum of the signal in (11) can be obtained, the spectral line after divided by $\alpha_1$ is

$$f_c + f_p : \gamma_1 = \sum_{m=1}^{M} m_m$$

$$f_c + 2f_p : \gamma_2 = \sum_{m=1}^{M} m_m \cdot e^{-j\beta D\sin\theta_m}$$

...

$$f_c + 2^{n-1}f_p : \gamma_n = \sum_{m=1}^{M} m_m \cdot e^{-j(n-1)\beta D\sin\theta_m}$$

It is obvious that this corresponds to the snapshot of the regular MUSIC algorithm. Dividing the corresponding spectral lines by $\alpha_3$ yields:

$$f_c + 3f_p : \gamma_1 = \sum_{m=1}^{M} m_m$$

$$f_c + 2 \cdot 3f_p : \gamma_2 = \sum_{m=1}^{M} m_m \cdot e^{-j\beta D\sin\theta_m}$$

...

$$f_c + 2^{n-1} \cdot 3f_p : \gamma_n = \sum_{m=1}^{M} m_m \cdot e^{-j(n-1)\beta D\sin\theta_m}$$

Similarly, we could also get this regular snapshot by other spectral lines.

Obviously, it is not enough to sample within a $T_p$. Therefore, we also need to perform similar operations within multiple $T_p$ to obtain more snapshots. Note that when two $T_p$ are too close to each other, the baseband signals corresponding to the incoming signals in both $T_p$ maybe very close, and the rank of $R_s$ cannot be improved. To ensure that the rank of $R_s = M$, we need to sample and perform frequency domain analysis within $M \times T_p$ at least.[3]

The snapshot vector obtained using this method has two distinct advantages over the conventional MUSIC method, as follows:

1. Since the Fourier transform has the meaning of averaging the noise, the SNR of the obtained snapshot from the frequency domain increases with the sampling frequency as well as the sampling duration.

2. Since we just need to analyze the frequency domain information of the combined signal, one RF channel is sufficient. This will greatly reduce the cost of equipment.

It should also be noted that the increase in the length of a single sample also brings disadvantages, the longer the sampling time the greater the fluctuations in the baseband signal, which will bring errors for the snapshot vector. Therefore, a reasonable trade-off is needed in the practical application.

The main calculation of the proposed method locates in the spectrum analysis to obtain the corresponding spectral line. Assume that $N_{pts}$ points' data are sampled in a $T_p$ and the total multiply-add operations for the frequency domain analysis are $N \times N_{pts}$ by the DFT and $N_{pts} \times N_{pts}$ by the FFT when only the first harmonics produced by each element is used.[4] Therefore, the calculation complexity of the proposed method is not high.

### III. SIMULATION RESULTS

To verify the feasibility and accuracy of this method, this section will use numerical simulations to verify it. A 4-element array time-modulated array is used in this simulation to orient three narrowband signals from the far field. First, after receiving the signals, the received signals are modulated using modulation functions on different array elements and then combined. This is followed by a frequency domain analysis. If not specified, we will use only the first harmonic generated by each array element for the directional measurements.

Expressions for the baseband signals of the three incoming signals:

$$s_{b(1)}(t) = 0.6i \times e^{i2\pi \cdot 7t}$$
$$s_{b(2)}(t) = -0.8i \times e^{i2\pi \cdot t}$$
$$s_{b(3)}(t) = 0.3i \times e^{i2\pi \cdot 4t}$$

The main simulation parameters are listed in Table I.

TABLE I
PARAMETERS FOR NUMERICAL SIMULATION

| Parameter | Values |
| --- | --- |
| Carrier frequency $F_c$ | 5000Hz |
| Element spacing $D$ | 30000m |
| Element number | 4 |
| Modulation frequency $F_p$ | 1250Hz |
| Number of $T_p$ | 5 |
| Sampling frequency $F_s$ | 16e5Hz |
| Total sampled points | $5 \times 1280$ |
| Signal-to-noise ratio (SNR) | 20 dB |

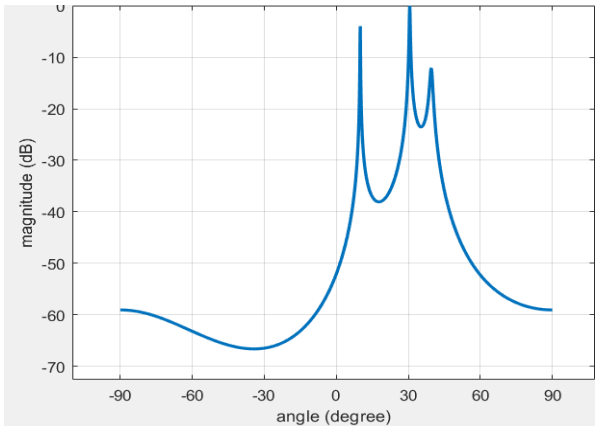

Figure 1. Spatial spectrum (θ1=10°, θ2=30°, θ3=40°)

## IV. CONCLUSION

In this paper, a novel direction-finding method of multi-narrowband signal based on a non-uniform periodic time-modulated array is described, in which the conventional snapshot can be obtained by analyzing the frequency domain information of the combined signal. The relevant formulas are derived in detail and the requirements for the incoming signal and the array when using this method are explained. In the last part of the paper, a numerical simulation is provided to verify the feasibility of this method. In future studies, the advantages, and disadvantages of the obtained snapshots in the frequency domain and some factors affecting the accuracy of the method such as the number of samples used for a frequency domain analysis, the incoming wave angle, the number of array elements, and the number of snapshots obtained in the frequency domain will be further discussed.